# FROM THE MAGNETIZATION PROFILE TO THE STRAY FIELD OF BISTABLE WIRES

P. Gawroński[1,*], A. Chizhik[2] and J. González[2]

[1]Faculty of Physics and Applied Computer Science, AGH University of Science and Technology, al. Mickiewicza 30, 30-059 Cracow, Poland

[2]Departamento Física de Materiales, Facultad de Química, UPV/EHU, 1072, 20080 San Sebastián, Spain

**Abstract**

We present new analytical calculations of the spatial dependence of the stray field of a bistable magnetic wire from the magnetization profile of the wire. Contributions from the outer shell and from the wire ends are neglected. The results qualitatively agree with experimental data, taken from literature.

## Introduction

Magnetic properties of bistable amorphous wires have attracted great interest not only due to their applications in sensors, MEMS technology and biomedical devices [1-3], but also from the micromagnetic point of view [4,5]. Amorphous wires are called bistable because of a single large Barkhausen jump which appears during the reentrant magnetization reversal process. This behavior is owing to a peculiar domain structure, determined by the stress frozen during the fabrication process. The structure includes an inner core, a single domain which could be axially magnetized up or down, and an outer shell [6]. Recent experimental data on Fe-rich wires [7] provide a new insight into a complicated maze-like surface structure of the outer shell. The authors of [7] suppose that Fe-rich wires possess unclosed 180º surface domain structure with magnetization perpendicular to the wire surface, without closure domains assumed previously for these materials.

Apart from the domain structure the magnetostatic interaction between two or more bistable amorphous wires is an interesting topic of research. The simplest possible model of the interaction is the dipolar model [8-10]. In this case the wire is replaced by two "magnetic charges" at the wire ends. This approximation has been analyzed carefully in [11] and criticized as too rough. Pure dipolar model is acceptable only if the distance between the wires is much larger than their length; in this regime, the interaction itself is negligible. Recent experimental data from the SQUID microscopic measurement of the stray magnetic filed profile [12] provide us, indirectly, with the information that the "magnetic charge" is not concentrated at the end of the wire but spread rather smoothly over the wire. An evaluation of an interaction between the wires should take this spreading into account.

The aim of this work is to present new calculations of the stray field of the FeSiB wire of diameter of 125 μm. This size and composition are often used to investigate the wire-wire interaction. The information we use as an input is the "magnetic charge" density, obtained from the measurement of the magnetization profile [13]. Next step is to calculate the magnetic scalar potential and, subsequently, the radial and the parallel component of the stray field. We compare the results of the calculations with accessible experimental data on the wire-wire interaction. In the following section we present the scheme of the calculation. Conclusions are given at the end of the text.

## Calculations and results

We consider two wires placed in parallel within the mutual distance *a* from almost zero to *2.5 mm* in an external magnetic field. By choice, zero of the z-axis is at the center of one of the wires. Each wire is a cylinder of the length of *L=10 cm*. The estimated value of the diameter of the inner core (i. e. 87 μm [14]) is used here as the wire diameter, since the wires are homogeneously magnetized along the wire axis only in the area of the inner core. In our preliminary approach we neglect the contribution of the outer shell to the stray field.

In order to proceed with our calculation we need a way to estimate the "magnetic charge" density. This information is provided by the measurement of the local magnetization profile, i.e. the position dependence of the magnetization along the wire [13]. Due to the limitation of the experimental technique, the measurement does not give a reliable value of the magnetization at the ends of the wire; therefore we assume that the magnetization at the end of the wire is equal to zero. The experimental data are fitted by the polynomial function of the 9[th] order, what gives us our starting point the *M(z)* dependence presented in Fig. 1. In our notation, positive values of the magnetization profile mean that the wire is magnetized "up" along the z-axis.

The so-called "magnetic charge" density is calculated [15] as

$$\rho = -\nabla \cdot M, \quad (1)$$

what in our case reduces to

$$\rho(z) = \frac{-\partial M(z)}{\partial z}. \quad (2)$$

The "magnetic charge" density $\rho(z)$ calculated according to Eq. 2 is presented in Fig. 2. Contrary to the dipole model the "magnetic charge" is not concentrated at the exact end of the wire but it is smeared smoothly across the whole wire. Following the electrostatic analogy, the magnetic scalar potential is defined as follows [15]

$$H_M = -\nabla \varphi_M. \quad (3)$$

The solution of the Poisson's equation applied to a magnetic body of finite dimension leads to [15]

$$\varphi_M(r) = \frac{1}{4\pi} \int_V \frac{\rho(r')}{|r-r'|} d^3 r' + \frac{1}{4\pi} \oint_S \frac{n \cdot M(r')}{|r-r'|} da', \quad (4)$$

where $|r-r'| = \sqrt{a^2 + (z-z')^2}$.

The second integral of Eq. 4 is performed over the wire boundary surface, and it is equal to zero as long as the magnetization at the wire ends is equal to zero. For the cylindrical coordinate system the parallel $(H_{Mz})$ and the radial $(H_{Mr})$ component of the stray field are calculated as $H_{Mz} = -\frac{\partial}{\partial z} \varphi_M(r)$ and $H_{Mr} = -\frac{\partial}{\partial r} \varphi_M(r)$. In our case the parallel $(H_{Mz})$ and the radial $(H_{Mr})$ component of the stray field are expressed as follows:

$$H_{Mz}(z) = \frac{R^2}{4\pi} \int_0^L \frac{\rho_M(z')(z-z')}{\left(a^2+(z-z')^2\right)^{3/2}} dz' \quad (5)$$

and

$$H_{Mr}(z) = \frac{R^2}{4\pi} \int_0^L \frac{\rho_M(z')}{\left(a^2+(z-z')^2\right)} dz' \quad (6)$$

We use Eqns. (5) and (6) to calculate the stray field produced by the bistable wire in the space around it. In Fig. 3a and 3b we present the radial component $(H_{mr})$ and the parallel component $(H_{Mz})$, respectively, of the stray field at the various distances from the wire. The calculated dependences can be divided into 3 regimes, namely; a) outside the wire, z is from -0.1 m to -0.07m and from 0.07m to 0.1m, b) at the wire ends around -0.05m and 0.05m, c) inside the wire, where z is from -0.05m to 0.05m. In the region (a) the stray field components are equal to zero. The values of the region (c) shown in Fig 3b are relevant for the interaction between parallel wires. The values obtained in region (b) have the same direction as the wire magnetization; therefore they cannot be responsible for the splitting of the hysteresis loop, known from the experiment [8,9].

**Discussion**

The calculated values of both stray field components at the wire ends do depend on our assumption made about the values of the magnetization at the end of the wire. Still, our data shown in Fig 3a could be compared with the stray field profile, shown in Ref. [12]. As follows from these measurements, the radial component $H_{Mr}$ gradually increases from zero to about 25 A/m, then monotonously decreases till zero near the wire center, decreases further to about -25 A/m and then tends to zero again. The same behavior is obtained in our calculations. To apply the method described in Section 2 the magnetization profile is needed of the wire of Ref. [12]; unfortunately this is not measured yet.

A typical axial hysteresis loop of two interacting wires consists of two Barkhausen jumps disjoint by a horizontal plateau [8]. The plateau length is equal to the interaction field multiplied by two. Let us consider the results concerning the interaction between the wires of nominal composition $Fe_{77.5}B_{15}Si_{7.5}$ found in recent papers. The strength of the interaction depends on the size of the wires and on their mutual distance. When two wires of nominal composition $Fe_{77.5}B_{15}Si_{7.5}$ and of diameter of 131 μm touch each other, the value of the interaction between them is about *5A/m* [8]. This value is in good agreement with our calculations, the strength of the interaction field taken from our Fig 3b is also about 5 A/m.

The scheme of calculation described above allows to calculate the interaction field dependence on the distance between the wires. However, there is an experimental evidence that the hysteresis loop of the interacting wires is not a direct measure of the stray field [16]. The plots 2a and 2b in Ref. [16] present the stress dependence of the interaction field for wires of the diameter of 125μm and 50μm, respectively. The stray field varies with the applied stress and this cannot be reduced to the variation of the magnetization. These results suggest that the position of the point where the switching process starts depends on the state of the wire. The stray field does depend on the position, and therefore it is not clear which value of the field is equivalent to the interaction between the wires. In Fig.4 we show the maximal value of the interaction field, as dependent on the distance *d* between wires. The obtained plot decreases with distance as $d^{-a}$, with *a=0.54*. This value of the obtained exponent is comparable to *a=0.49* obtained when fitting the data of Ref. [17] for Fe-rich microwires. However, in the latter case the fitting is rather poor. More generally *i)* we do not see any reason why should the proportionality of the stray field to $d^{-a}$ be valid; we treat the value $x$ only as a determinant of the character of the data, *ii)* we are not convinced why the switching field is connected to the maximal value of the stray field. If, on the contrary, we calculate the stray field value at a given point at the wire, the curve character shown in Fig. 5 is qualitatively similar to the experimental data of Ref. [17]. The difference in values is due to the difference in the wire size. The character of the curve remains the same.

Concluding, we presented a theoretical scheme how the data on magnetization profile of a bistable wire can be used to calculate the spatial dependence of the stray field around the wire. Obtained values of the stray field agree qualitatively with the experimental data, found in literature. Still observed discrepancies are due, in our opinion, at least partially to the fact that the measurement of the magnetization profile does not allow to determine accurately the density of the magnetic charge at the wire ends. Further calculations should take into account also the screening of the stray field by the outer shell.

## Acknowledgment

One of the authors (P. G.) wishes to thank Krzysztof Kułakowski for critical reading of the manuscript and helpful discussions, and Arcady Zhukov for supplying the experimental data for Fig. 1. This work was supported by MEyC under project PCI2005-A7-0230.

## References

[1] R. J. von Gutfeld, J. F. Dicello, S. J. McAllister and J. F. Ziegler, Appl. Phys. Letters 81 (2002) 1913.

[2] M. Barbic, J. Magn. Magn. Mater. 249 (2002) 357.

[3] A. Zhukov and J. González, in *Encyclopedia of Sensors*, Vol. X, C. A. Grimes, E. C. Dickey and M. V. Pishko (Eds.), American Scientific Publishers, 2006, pp. 1-25.

[4] V. Zhukova, N. A. Usov, A. Zhukov and J. González, Phys. Rev. B 65 (2002) 134407.

[5] R. Hasegawa, J. Non-Cryst. Solids 329 (2003) 1.

[6] F. B. Humphrey, K. Mohri, J. Yamasaki, H. Kawamura, R. Malmhall and J. Ogasawara, in *Magnetic Properties of Amorphous Metals*, edited by A. Hernando et al. (Elsevier, Amsterdam 1987).

[7] Yu. Kabanov, A. Zhukov, V. Zhukova and J. González, Appl. Phys. Lett. 87 (2005) 142507.

[8] J. Velazquez, C. Garcia, M. Vazquez and A. Hernando, Phys. Rev. B 54 (1996) 9903.

[9] L. C. Sampaio, E. H. C. P. Sinnecker, G. R. C. Cernicchiaro, M. Knobel, M. Vazquez and J. Velazquez, Phys. Rev. B 61 (2000) 8976.

[10] J. Velazquez, C. Garcia, M. Vazquez and A. Hernando, J. Appl. Phys. 81 (1997) 5725.

[11] J. Velázquez, K. R. Pirota and M. Vazquez, IEEE Trans. Magn. 39 (2003) 3049.

[12] S. Gudoshnikov, N. A. Usov, A. Zhukov, J. González and P. Palvanov, Paper I-Ep-012, presented at the III Joint European Magnetic Symposia, San Sebastian, Spain, 26-30 June 2006; J. Magn. Magn. Mater. (2007), in print.

[13] V. Zhukova, A. Zhukov, J. M. Blanco, J. González, C. Gomez- Polo and M. Vazquez J. Appl. Phys., 93 (2003) 7208.

[14] F. Kinoshita, IEEE Trans. Magn. 26 (1990) 1786.

[15] G. Bertotti, *Hysteresis in Magnetism*, Academic Press, San Diego, 1998.

[16] P. Gawroński, A. Zhukov, J. M. Blanco, J. González and K. Kułakowski, J. Magn. Magn. Matter. 290-291 (2005) 595.

[17] A. Chizhik, A. Zhukov, J. M. Blanco, R. Szymczak and J. González, J. Magn. Magn. Mater. 249 (2002) 99.

**Captions**

Fig. 1. The magnetization profile of the Fe-rich bistable wire [9]. Both experimental points and the fitted curve are shown.

Fig. 2. The density of the magnetic charge, as calculated from the data shown in Fig. 1.

Fig. 3. The stray field dependence on the position *z* along the wire, in different distances: *a)* the radial coordinate, and *b)* the coordinate parallel to the wire . The distance from the wire is 0.5 mm, 1 mm, 2 mm and 5 mm for the presented curves.

Fig. 4. The distance dependence of the maximum of the interaction field, shown in Fig. 3b.

Fig. 5. The distance dependence of the interaction field, i.e. the stray field component parallel to the wire, at the distance 5 mm from the wire end ($H_U$ =1A/m, $z_U$ = 1m). The calculated curve (squares) and the experimental data (circles) are of the same character.

Fig. 1

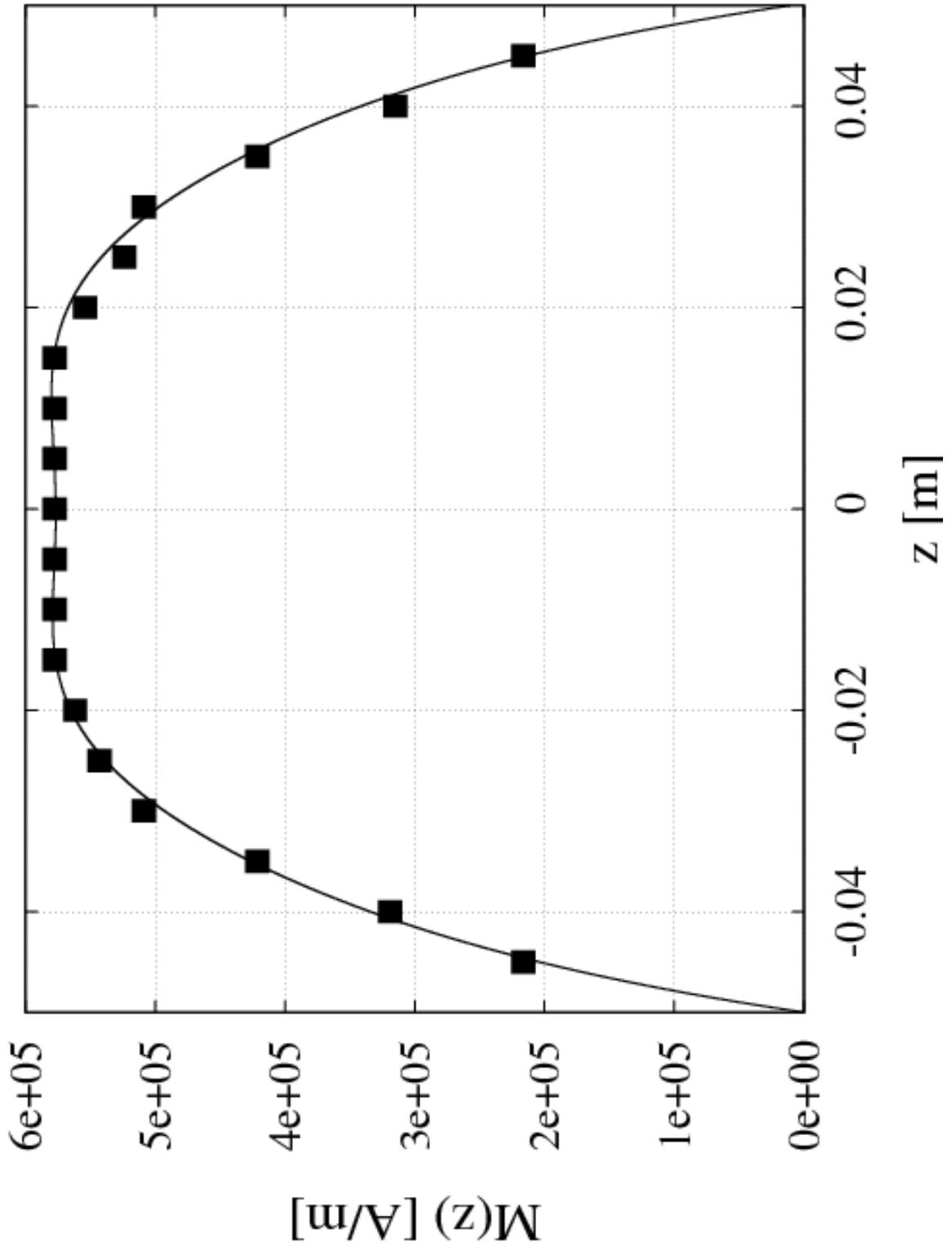

Fig. 2

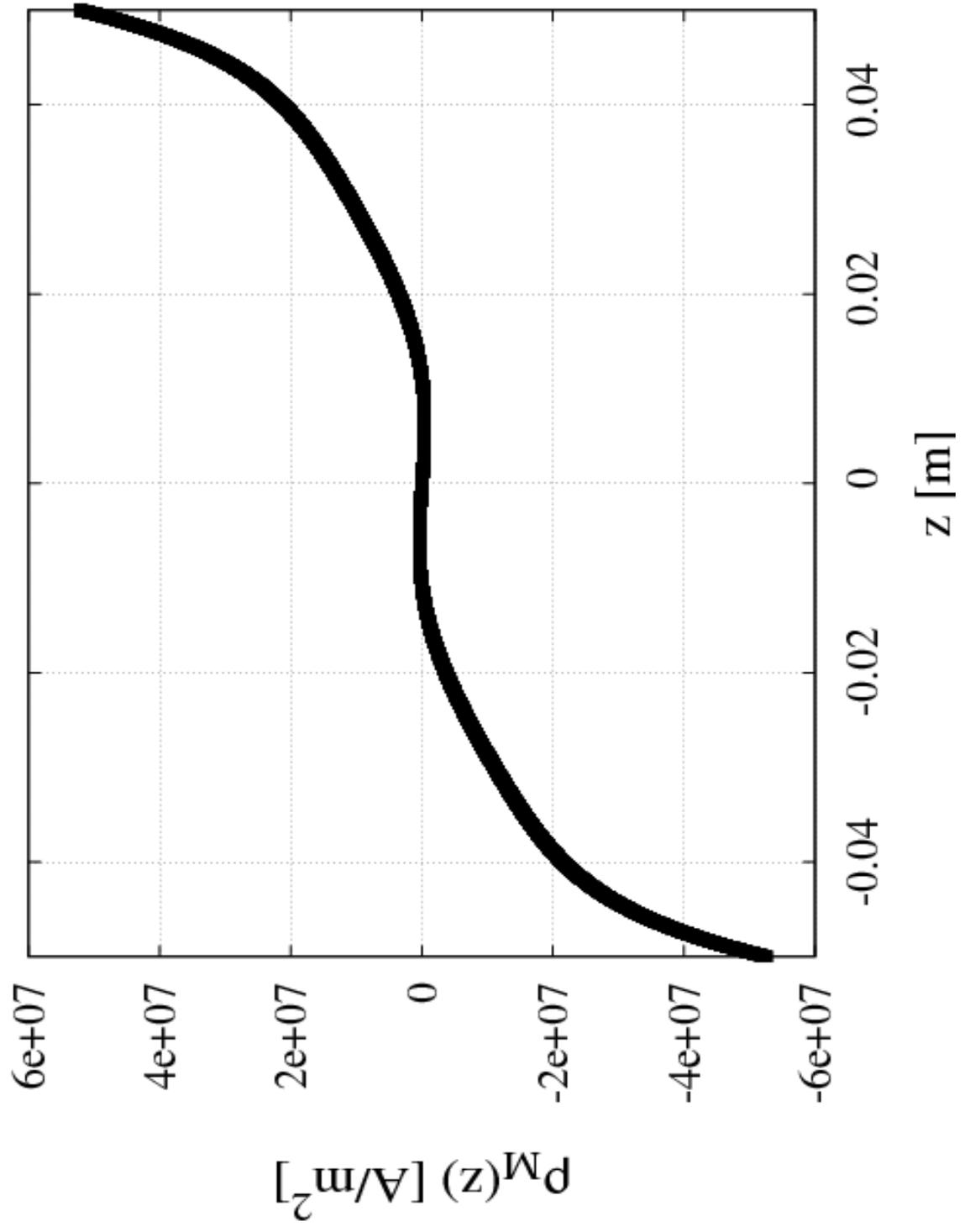

Fig. 3 a

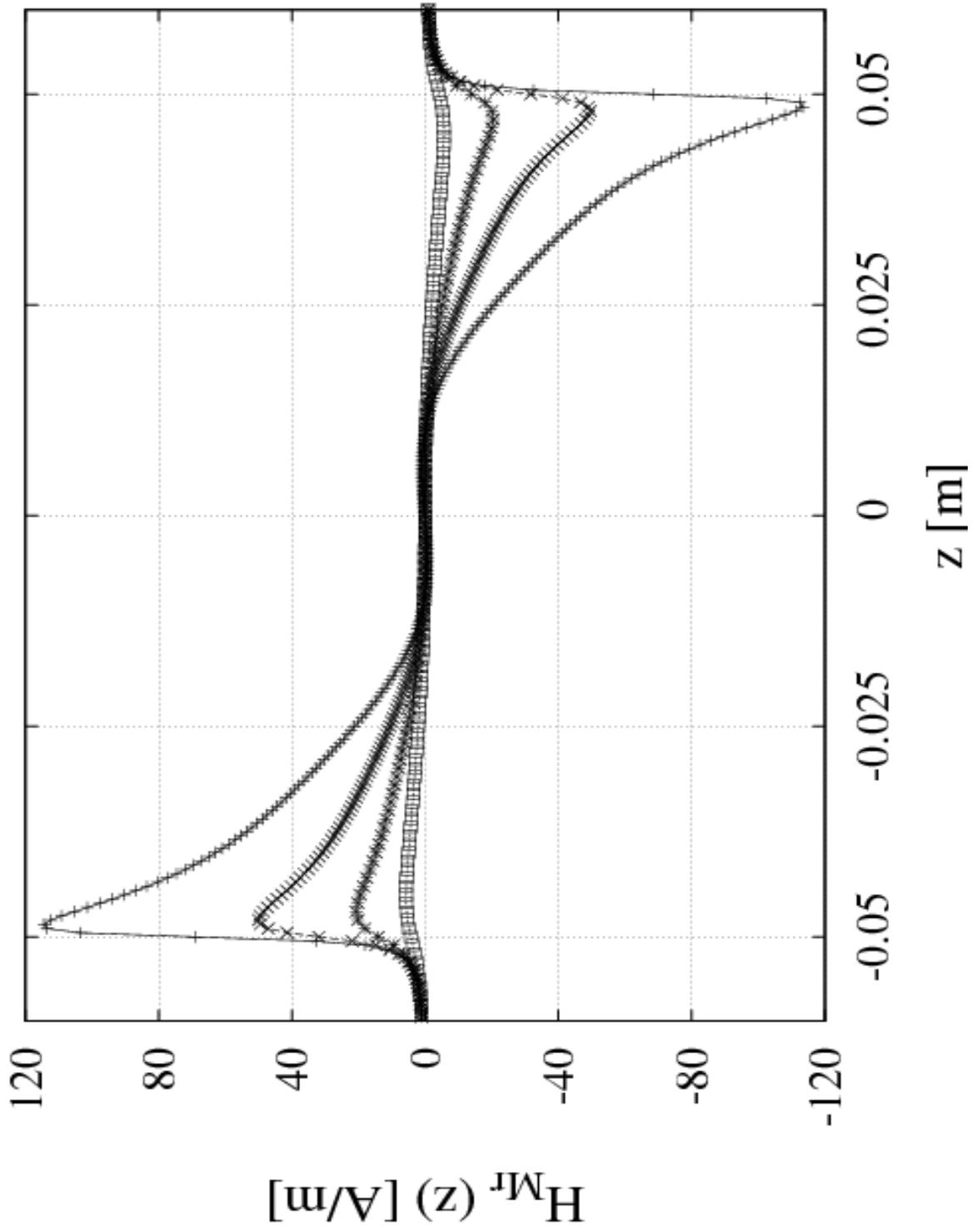

Fig. 3 b

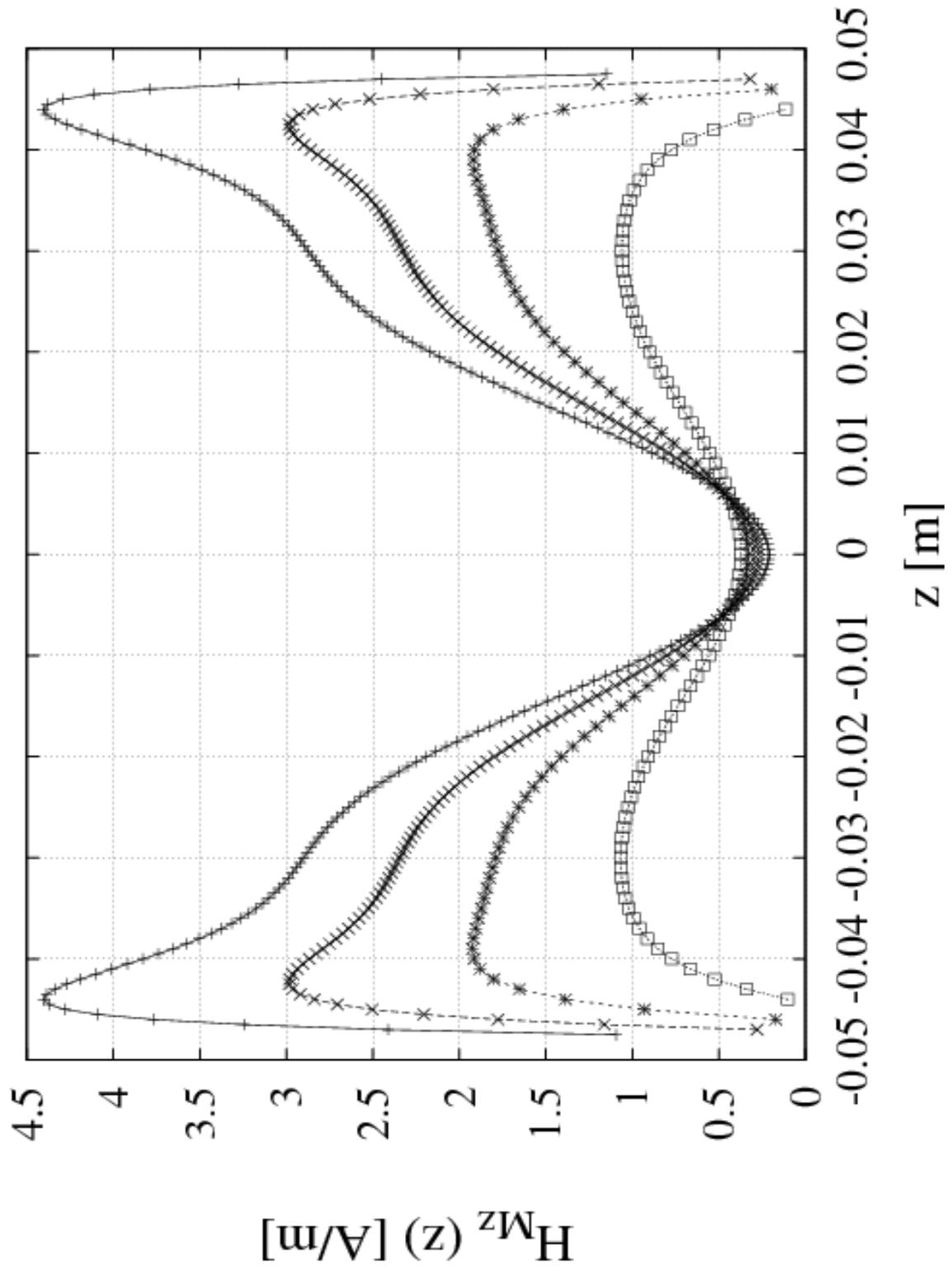

Fig 4

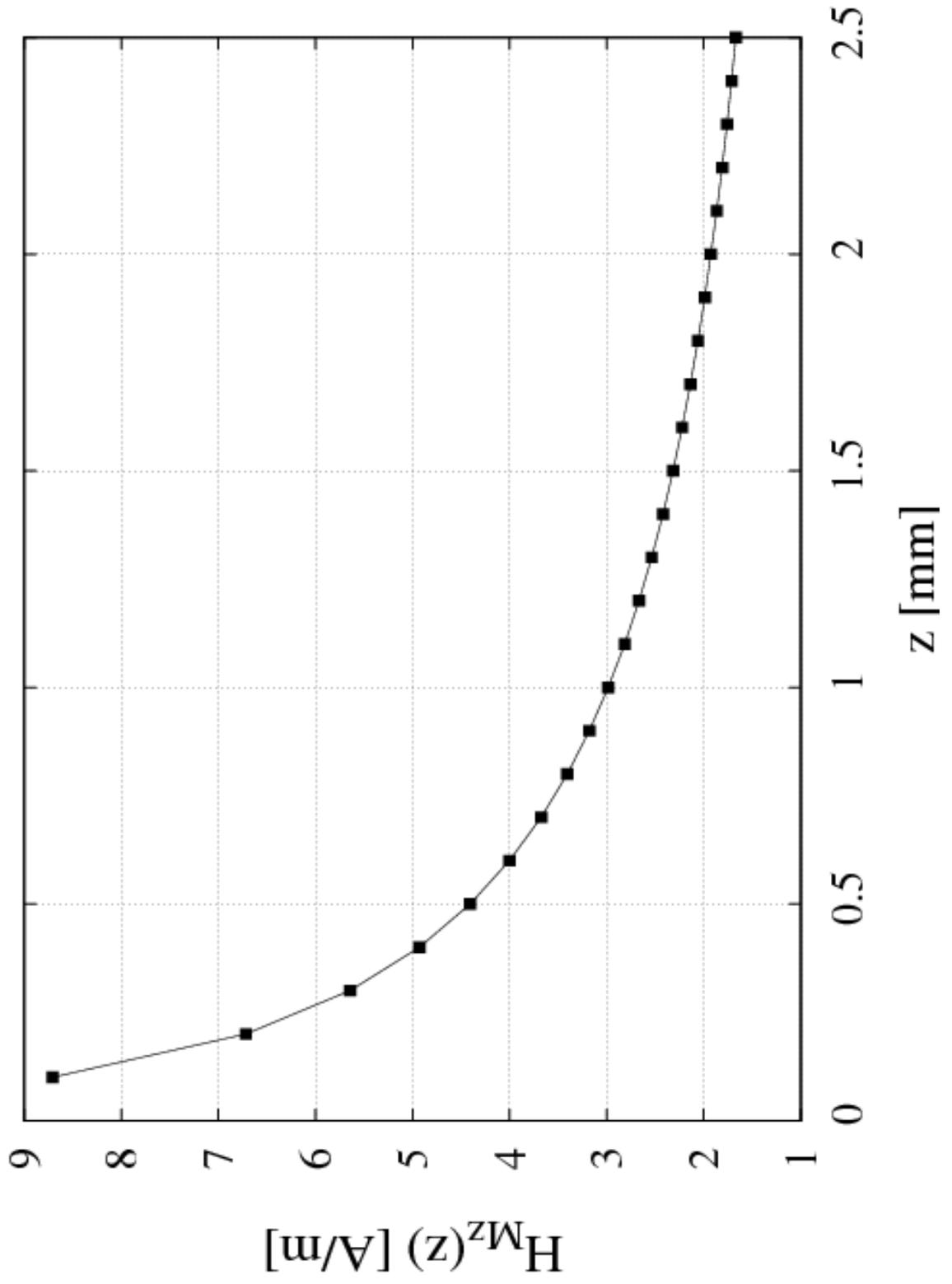

Fig. 5

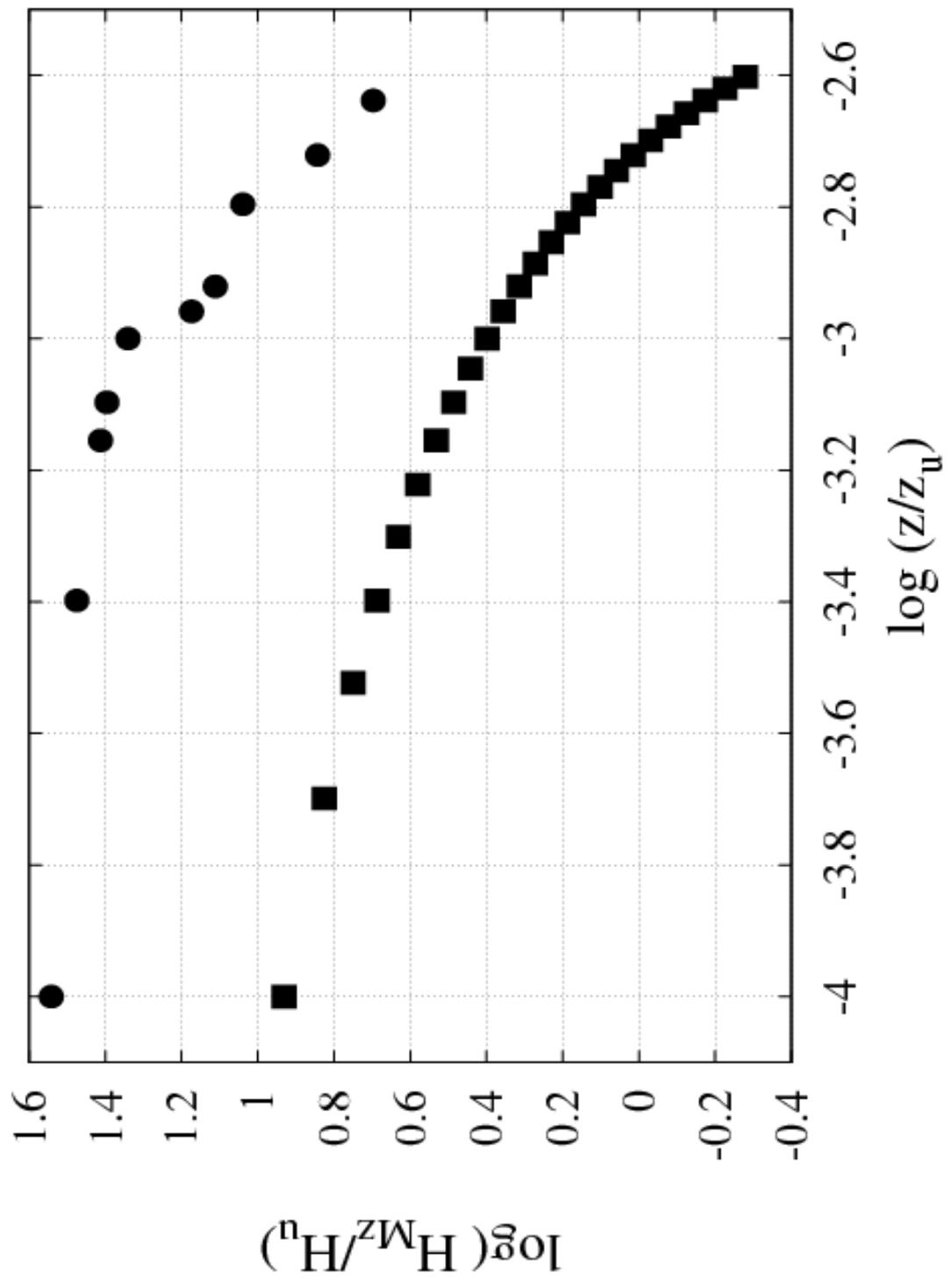